\documentclass[letterpaper,10pt]{article}
\usepackage[utf8]{inputenc}
\usepackage[T1]{fontenc}
\usepackage{kpfonts}
\usepackage{textcomp}

\usepackage{graphicx}
\usepackage[english]{babel}
\usepackage[table]{xcolor}
\usepackage{amsmath,amsfonts,amssymb}
\usepackage[version=3]{mhchem}
\usepackage{array}

\usepackage{fullpage}
\usepackage{subfig}
\usepackage{booktabs}								
\usepackage{multirow}	
\usepackage[section]{placeins}
\usepackage{pdflscape}
\usepackage{tabularx}
\usepackage{natbib}
\usepackage{fancyref}
\usepackage{authblk}
\usepackage{longtable}

\usepackage{setspace}
\usepackage{lineno}

\begin{document}

\title{Experimental study and reaction path modeling of the carbonation of natural serpentinites \\F. Osselin, M. Pichavant, A. Lassin \\
florian.osselin@cnrs-orleans.fr}

\maketitle 


\section*{Abstract}
Mineralization of carbon dioxide is often seen as an attractive alternative to classical Carbon Capture and Storage (CCS) technologies, allowing the sequestration of \ce{CO2} as a solid mineral with no risk of aquifer contamination or leakage back to the atmosphere. While olivine and pyroxenes are known to easily and quickly react with dissolved \ce{CO2}, fresh peridotites are quite rare and ultramafic rocks usually contain significant amounts of serpentine, which presents a lower reactivity. The purpose of this study was then to analyze the reactivity of two natural rocks: a partially serpentinized lherzolite and a fully altered serpentinite. Results confirm that serpentine is much slower to react and gets altered only if the activity of \ce{CO2} is high enough and if all olivine and pyroxenes have already been consumed. Resulting carbonates are mostly Mg-rich calcite or Mg-depleted dolomite with the occurrence of eitelite (\ce{Na2Mg(CO3)2} in the case of high Na activities. The carbonation of these serpentinite was however associated in some cases with a heavy precipitation of hazardous asbestiform chrysotile, which could be a potential threat for engineered carbonation processes. 

\section{Introduction}
Among the different techniques currently under development for the geological sequestration of carbon dioxide, \ce{CO2} mineralization presents the obvious advantage of producing stable and solid materials \citep{Matter2009}. This is in contrast with the classic CCS  (Carbon Capture and Storage) view of injecting carbon dioxide in porous formations. While the latter takes advantage of the immense storage capacities of deep saline aquifers \citep{Bachu2003}, depleted oil and gas fields or salt cavities, it still presents the drawback of requiring constant monitoring during and, most importantly, after the injection, to make sure the \ce{CO2} does not leak back to the atmosphere or disturb shallow potable aquifers through permeable formations, faults or improper casing sealing \citep{Deng2017,Vinca2018}. Conversely, carbon mineralization, takes advantage of the chemical reactivity of the mineral feedstock to precipitate directly solid carbonates such as magnesite $\ce{Mg^{2+} + HCO_3^- = MgCO3(s) + H^+}$, which is stable over a very large range of pressure and temperatures and can even be of economical value. In order to react with the \ce{CO2}, the feedstock is first dissolved in water (sometimes through a pretreatment stage, e.g. \citet{Krevor2011}) and releases divalent cations (i.e. $\ce{Mg^{2+}}$, $\ce{Ca^{2+}}$\ldots), which combine with the dissolved carbon dioxide to form solid carbonates. 

Among the different targets for carbon mineralization, the most studied type of minerals are silicate, and primarily mafic and ultramafic rocks \citep{Mcgrail2006,Oelkers2008,Kelemen2011} containing olivine \ce{Mg2SiO4}, ortho- and clinopyroxenes (\ce{MgSiO3}, \ce{MgCaSi2O6}) as well as the rarer wollastonite \ce{CaSiO3} \citep{Daval2009}. As a result, a wealth of publications have studied the influence of temperature, pressure, solution and feedstock composition on the yield and kinetics of these reactions (e.g. \citep{Alexander2007,Lafay2018,Park2003,Lisabeth2017a,Hovelmann2012,Orlando2011,Truche2021}. Results show that the fastest reaction and highest carbonation rates were consistently obtained with wollastonite closely followed by olivine. On the contrary, serpentine \ce{Mg3Si2O5(OH)2} was found  to be significantly slower  (despite some experiments suggesting the opposite  at 70\textcelsius\ (\citep{Lacinska2017}). The optimal conditions for carbonation of olivine were determined in the extensive research carried out in the Albany Research Center \citep{Oconnor2005} as 1M \ce{NaCl}, 0.64M {NaHCO3} at 185\textcelsius\ and 150 bar of \ce{CO2} partial pressure. However, the vast majority of these studies focused on pure olivine and pure minerals, and very few considered natural rocks (see for example \citet{Hovelmann2011}) despite their closer proximity to actual mineralization processes. In particular, serpentinites, rocks derived from the serpentinization of peridotites and containing both primary ultramafic minerals (olivine, pyroxenes) as well as serpentine and other secondary phases (magnetite, aragonite, talc\ldots), represent a very interesting target for \ce{CO2} mineralization projects due to their  better accessibility and larger available volume than fresh peridotites. Moreover, serpentinites are also currently used for nickel extraction, while its asbestos form (chrysotile) has extensively been mined for decades before its interdiction in most countries due to serious health issues. Using these wastes as feedstock offers an ideal solution for ex-situ mineral carbonation as well as a potential solution for asbostos inertization as the process would destroy the chrysotile structure and thus its asbestiform characteristics. 

In this study, we explore the reactivity of two natural serpentinites, one dredged from the South-West Indian Ocean Ridge (SWIR) and the other obtained from the Ocean Drilling Project (ODP leg 209). The purpose is to explore the stability of these serpentinites and, in particular, their capacities for carbon dioxide mineralization without any pretreatment. To do so, a series of experiments in gold capsules was conducted at 165\textcelsius\ and 300\textcelsius\ and at 500 bars. The resulting mineral phases were analyzed by X-ray diffraction (XRD) and Scanning Electron Microscopy coupled with Electron Dispersive Spectroscopy (SEM-EDS) to characterize the reaction progress and the reactivity of the starting material under the different conditions.

\section{Materials and Methods}

\subsection{Starting Materials}
Most of the experiments were proceeded with a natural serpentinite dredged from the South-West Indian Ocean ultraslow Ridge (SWIR) \citep{Roumejon2014}. The serpentinite was characterized by XRD, SEM/EDS, electron-microprobe analysis (EPMA) and Raman spectroscopy. The modal composition, obtained from point counting on 3 different thin sections (0.5mm$\times$1mm step) is variable across the protolith but in average  represents a partially serpentinized lherzolite  ($\approx$50\%wt serpentine) containing relicts of olivine, ortho- and clinopyroxenes. The serpentine polytype was also identified by Raman spectroscopy as principally lizardite, with a few chrysotile veins. This serpentinite also underwent several carbonation events leading to a pervasive network of carbonate veins -$\approx$4-5\%wt- identified on Raman spectra as aragonite.

The second starting material used in this study is a serpentinite from the Ocean Drilling Project leg 209 \citep{Paulick2006} hole 1268, Core 19 Section 1 (plugged between the 20 and 30cm marks). This serpentinite is composed exclusively of serpentine with minor magnetite and NiFe sulfate, and traces of chromite. The thin section reveals characteristic mesh (after olivine) and bastite (after pyroxene) textures. Similarly to SWIR, the serpentine is lizardite with some chrysotile veins. 

\begin{figure}
	\centering \includegraphics[width=\textwidth]{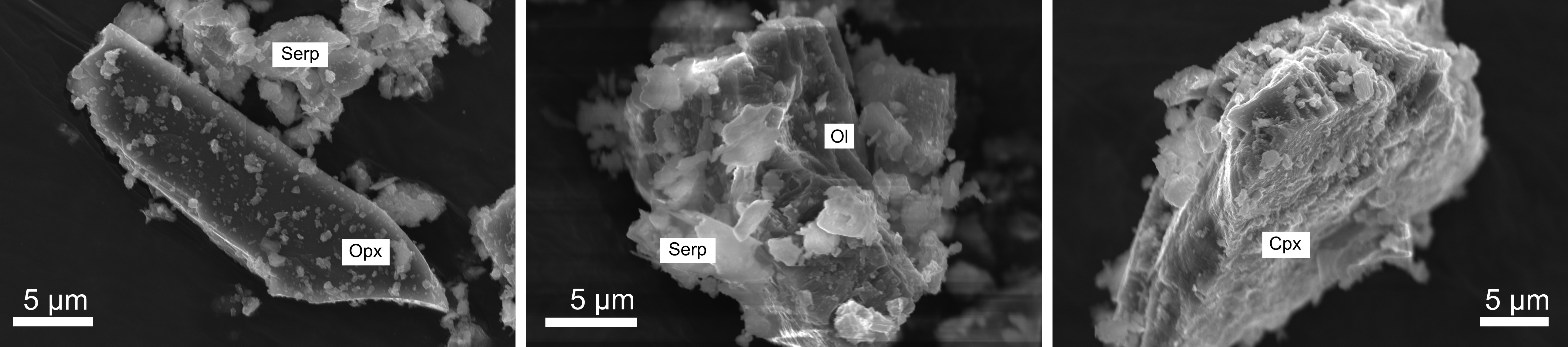}
	\caption{\label{fig:SerpCan_powder} SEM images from the SWIR serpentinite. From left to right, an orthopyroxene (opx) fragment, an olivine grain with serpentine flakes (lizardite) and finally a clinopyroxene (cpx) grain}
\end{figure}

Table \ref{tab:composition-protoliths} gives the whole rock composition determined by ICP-OES for major elements and ICP-MS for traces at SARM (Service d'Analyses des Roches et des Minéraux) as well as the principal constituting minerals composition from EPMA.

\begin{table}
\begin{tiny}
	\centering
	\begin{tabular}{ccccccccccccccccccccc}
		\toprule
\multicolumn{2}{c}{\textbf{Whole Rock}} & \ce{SiO2} & \ce{MgO} & \ce{CaO} & \ce{Fe2O3_{tot}} & \ce{Al2O3}  & \ce{Na2O} & \ce{MnO} & \ce{K2O}  & \ce{TiO2} & \ce{H2O} & \ce{CO2_{tot}} & & \ce{Co} & \ce{Cu} & Cr  & Ni& Sr & V & Zn  \\
 \multicolumn{2}{c}{\textbf{Analysis}} &  \multicolumn{11}{c}{\%wt} & & \multicolumn{7}{c}{ppm} \\
\cmidrule{1-1} \cmidrule{3-13} \cmidrule{15-21}
SWIR  & & 41.41 & 33.31 & 5.03 & 8.29 & 3.07 & 0.15 & 0.11 & < lq & 0.091 & 6.38 & 1.88 & & 90.0 & 22.1 & 2567  & 1703 & 403 & 75.7 & 48.2 \\
\cmidrule{1-1} \cmidrule{3-13} \cmidrule{15-21}
ODP-1268-19R1 & & 37.79 & 38.59 & < lq & 7.44 &  0.45 & 0.07 & 0.067 & < lq &  < lq & 13.36 & 0.21 & & 95.6 & 22.9 &  1856 & 1942  & 1.2 & 19.3 & 28.2 \\
\midrule[1pt]
\multicolumn{2}{c}{\textbf{EPMA}} & & & & & & & & & & & & & & & & & & & \\
\cmidrule{1-1} 
\multicolumn{2}{c}{Olivine-SWIR} & 40.0 & 48.7 & 0.05 & 10.6 & 0.02 & 0.05 & 0.13 & 0.01 & 0.03 & n.q. & n.q. & & & & & & & & \\
\cmidrule{1-1} \cmidrule{3-13} 
\multicolumn{2}{c}{Opx-SWIR} & 52.4 & 31.8 & 1.19 & 6.63 & 5.72 & 0.06 & 0.13 & 0.02 & 0.06 & n.q. & n.q. & & & & & & & & \\
\cmidrule{1-1} \cmidrule{3-13} 
\multicolumn{2}{c}{Cpx-SWIR} & 50.5 & 15.1 & 22.7 & 2.83 & 6.93 & 0.75 & 0.14 & 0.01 & 0.48 & n.q. & n.q. & & & & & & & & \\
\cmidrule{1-1} \cmidrule{3-13}
\multicolumn{2}{c}{Serpentine (mesh)} & 39.4 & 37.7 & 0.07 & 6.24 & 0.2 & 0.02 & 0.07 & 0.01 & 0.04  &n.q. & n.q. & & & & & & & & \\
\cmidrule{1-1} \cmidrule{3-13} 
\multicolumn{2}{c}{Serpentine (bastite)} & 38.5 & 32.7 & 0.8 & 8.0 & 5.5 & 0.01 & 0.13 & 0.01 & 0.00 & n.q. & n.q. & & & & & & & & \\
\cmidrule[1pt]{1-13}
\multicolumn{2}{c}{\textbf{EDS - norm. 100\%}} & & & & & & & & & & & & & & & & & & & \\
\cmidrule{1-1}
\multicolumn{2}{c}{Serpentine OPD} & 49 & 45 & 0 & 2.5 & 0.5 & 0 & 0 & 0 & 0 & n.q. & n.q. & & & & & & & & \\
\cmidrule[1pt]{1-13} 
\end{tabular}

\end{tiny}
\caption{\label{tab:composition-protoliths} Whole rock and EPMA analysis of the two protoliths used in the experiments. <lq = below limit of quantification, n.q. = not quantified}
\end{table}

\subsection{Experimental procedure}
A total of 14 high temperature-high pressure experiments were performed with durations up to 2 months. All experiments were proceeded with rapid quenching cold-seal autoclaves. Their characteristics are reported on Table \ref{tab:summary_exp}. The starting material (ground and sieved to keep the fraction below 50$\mu$m) as well as the different reactants were loaded in a gold tube welded shut to create a capsule 5 mm diameter and 3-4 cm long. 

Experiments SWIR300, ODP300 and SWIR165 were performed with a 2\% NaCl (Alfa Aesar 99+\% purity), 5\% \ce{NaHCO3} (SigmaUltra 99.5\% purity) at 165\textcelsius\ and 300\textcelsius (i.e. 0.34M NaCl and 0.6M \ce{NaHCO3}). Room temperature pH of this solution was measured as around 8. As results of these experiments revealed that serpentine is only destabilized after all the primary minerals are consumed and if the \ce{CO2} activity is high enough, other fluid compositions, using silver oxalate (\ce{Ag2C2O4} - American Elements), sodium oxalate (\ce{Na2C2O4} - Sigma Aldritch  >99.5\% purity) or simply an excess of \ce{NaHCO3}, were designed in order to analyze the effect of higher \ce{CO2} activities on the system. The total C over the total cations from the primary minerals (olivine, orthopyroxenes and clinopyroxenes) is reported on Table \ref{tab:summary_exp}.

Reactants were either inserted as an already prepared solution for the initial solution, or as powder for \ce{Ag2C2O4} and \ce{Na2C2O4}. The water used was deionized water. For both SWIR300g experiments (see Table \ref{tab:summary_exp}), no welding was made and two porous ceramics were used instead to close the gold tube. 

For each experiment, except for the two SWIR300g, the autoclave was pressured with Ar gas up to 500 bars and heated to the desired temperature. In the case of the two SWIR300g experiments, the autoclave was pressured with \ce{CO2} instead. Temperature was controlled by a sheathed type K thermocouple inserted in the hot part. Once the experiment was finished, capsules were quenched, retrieved and cut open to recover the resulting solids. No fluid nor gas analysis were performed. Solids were then rinsed with deionized water, dried overnight in an oven at 120\textcelsius\ and analyzed by XRD, and SEM coupled with EDS. 

\begin{landscape}
\begin{tiny}
\begin{longtable}{llcccccccccccccl}

\toprule
 	Experiment  	& Starting	& Temperature	& Duration & \multirow{2}{*}{Capsule} & Powder & Water & W/R & \multirow{2}{*}{Fluid composition} & NaCl & NaHCO3 & Ag2C2O4 & Na2C2O4 & molC/cations & \multirow{2}{*}{Final Color} & \multirow{2}{*}{Comments} \\
 	\cmidrule{3-4} \cmidrule{6-8} \cmidrule{10-13}
 	name			& Material	& \textcelsius\	& Days &  & mg & mg & - &  & mg & mg & mg & mg& (ol+opx+cpx) & & \endhead
 	\midrule[1pt]
 	SWIR300g-short & SWIR & 300 & 6 & Au & 250 & 0 & - & \ce{CO2}$_{sc}$ & - & - & - & - & - & unchanged & 500 bars dry p$_{\ce{CO2}}$ \\
 	\midrule 
 	SWIR300g-long & SWIR & 300 & 33 & Au & 250 & 0 & - & \ce{CO2}$_{sc}$ & - & - & - & - & - & unchanged & 500 bars dry p$_{\ce{CO2}}$ \\ 	\midrule[1pt]
 	SWIR300-short & SWIR & 300 & 7 & Au & 48.6 & 105.5 & 2.2 & \ce{H2O}-2\%NaCl-5\%\ce{NaHCO3} & 2.1 & 5.27 & 0 & 0 & 0.31 & Red & Unidentified sandrose mineral \\
 	\midrule
 	SWIR300-medium & SWIR & 300 & 13 & Au & 49.7 & 92.9 & 2.2 & \ce{H2O}-2\%NaCl-5\%\ce{NaHCO3} & 1.86 & 4.64 & 0 & 0 & 0.27 & Red & \\
 	\midrule
 	SWIR300-long &  SWIR & 330 & 29 & Au & 49.8 & 98.1 & 2.0 & \ce{H2O}-2\%NaCl-5\%\ce{NaHCO3} & 2.7 & 9.2 & 0 & 0 & 0.53 & Red & \\
 	\midrule[1pt]
 	SWIR300-excess & SWIR & 301 & 31 & Au & 46 & 96.9 & 2.1 & \ce{H2O}-\ce{NaHCO3} in excess & 0 & 52.3 & 0 & 0 & 3.3 & Green \\
 	\midrule[1pt] 
 	ODP300 & ODP & 302 & 31 & Au & 49.5  &104.3 & 2.1 & \ce{H2O}-2\%NaCl-5\%\ce{NaHCO3} & 2.09 & 5.4 & 0 & 0 & - & Green & No identified carbonate \\
 	\midrule[1pt] 
 	SWIR165-short & SWIR & 165 & 13 & Au & 47.1 & 113.6 &  2.4 & \ce{H2O}-2\%NaCl-5\%\ce{NaHCO3} & 2.27 & 5.68 & 0 & 0 & 0.35 & Yellow & \\
 	\midrule
 	SWIR165-medium & SWIR & 163 & 21 & Au & 51.1 & 108.6 & 2.4 & \ce{H2O}-2\%NaCl-5\%\ce{NaHCO3} & 2.17 & 5.43 & 0 & 0 & 0.31 & Yellow & Unidentified net-like structure \\
 	\midrule
 	SWIR165-long & SWIR & 170 & 63 & Au & 50.3 & 82.5 & 1.6 & \ce{H2O}-2\%NaCl-5\%\ce{NaHCO3} & 4.65 & 4.1 & 0 & 0 & 0.24 & Red & Unindentified sandrose and net-like minerals\\
 	\midrule[1pt]
  	SWIR-Agox & SWIR & 297 & 30 & Au & 22.2 & 15.6 & 0.7 & \ce{H2O}-\ce{NaCl}-\ce{NaHCO3}-\ce{Ag2C2O4} & 0.31 & 0.78 & 117.2 & 0 & 8.1& Red \\
 	\midrule
 	ODP-Agox & ODP & 301 & 31 & Au & 30.1 & 15.4 & 0.51 & \ce{H2O}-\ce{NaCl}-\ce{NaHCO3}-\ce{Ag2C2O4}  & 0.31 & 0.76 & 112.8 & 0 & - & Red \\ 	
 	\midrule[1pt]
 	SWIR-Naox  & SWIR & 300 & 28 & Au & 29.2 & 13.9 & 0.48 & \ce{H2O}-\ce{Na2C2O4} & 0 & 0 & 0 & 46.4 & 5.5 & Green & unidentified sandrose mineral \\
 	\midrule
 	OPD-Naox & OPD & 301 & 31 & Au & 16 & 19 & 0.84 & \ce{H2O}-\ce{NaCl}-\ce{NaHCO3}-\ce{Na2C2O4} & 0.32 & 0.8 & 0 & 45.2 & - & Green \\
 	\bottomrule
 	
 	\caption{\label{tab:summary_exp}Characteristics of the different experiments presented in this study - W/R = water/rock mass ratio)}
\end{longtable}

\end{tiny}
\end{landscape}

\begin{landscape}
\begin{tiny}
\begin{longtable}{lcccccccccccccccp{1.7cm}}

\toprule
	\multirow{2}{*}{Experiment} & \multirow{2}{*}{Olivine} & \multirow{2}{*}{Opx} & \multirow{2}{*}{Cpx} & \multicolumn{2}{c}{Serpentine} & \multirow{2}{*}{Talc} & \multirow{2}{*}{Amphibole} &  \multirow{2}{*}{Magnetite} & \multirow{2}{*}{Hematite} & \multirow{2}{*}{Aragonite} & \multirow{2}{*}{Magnesite} & \multirow{2}{*}{Calcite} & \multirow{2}{*}{Dolomite} & \multirow{2}{*}{Eitelite} & \multirow{2}{*}{Clay} & \multirow{2}{*}{Others} \\
	\cmidrule{5-6} 
	& &	&	& Generic & fibrous (SEM) & &&&&&&&&&& \\
	\midrule
	SWIR & XRD/SEM & XRD/SEM & XRD/SEM & XRD/SEM & traces & -/- & -/- & XRD/SEM & -/- & XRD/SEM & -/- & -/- & -/- & -/-   & -/- & Chromite \\
	\midrule 
	OPD & -/- & -/- & -/- & XRD/SEM & traces & -/- & -/- & XRD/SEM & -/- & -/- & -/- & -/- & -/- & -/- & -/-  & traces of NiFe sulfide and Chromite \\
	\midrule[1pt]
	SWIR300g-short & XRD/NA & XRD/NA & XRD/NA & XRD/NA & NA & -/NA & -/NA & XRD/NA & -/NA & XRD/NA & -/NA & -/NA &-/NA &-/NA & -/NA & - \\
	\midrule 
	SWIR300g-long & XRD/NA & XRD/NA & XRD/NA & XRD/NA & NA & -/NA & -/NA&  XRD/NA & -/NA & XRD/NA & -/NA & -/NA &-/NA &-/NA & -/NA & - \\
	\midrule[1pt]
	SWIR300-short & XRD/- & -/- & -/- & XRD/SEM & - & -/- & -/- & -/- & XRD/SEM(?) & -/- & -/- & XRD/SEM\footnote{0-10\%wtMgO}  & -/-  & -/-  & XRD/SEM(?)  & \\
	\midrule
	SWIR300-medium & XRD/- & -/- & -/SEM & XRD/SEM & - & -/- & -/- &  -/- & XRD/- & -/- & -/- & XRD/SEM & -/- & -/- &  XRD/SEM  &  \\
	\midrule
	SWIR300-long & -/- & -/- & -/- & XRD/SEM & - & -/- & -/- & -/- & XRD/SEM(?) & -/- & -/- & XRD/SEM\footnote{0-5\%wtMgO} &  XRD/SEM & -/- &  XRD/-  & \\
	\midrule[1pt]
	SWIR300-excess & -/- & -/- & XRD/- & XRD/SEM & major & -/- & XRD/SEM & XRD/- & -/- & -/- & -/- & -/- & XRD\footnote{some \%wt FeO}/SEM - & XRD/SEM & XRD/- & Mg-Arfvedsonite, CSH \\
	\midrule[1pt]
	ODP300 & -/- & -/- & -/- & XRD/SEM & traces & -/- & -/- & XRD/- & -/- & -/- & -/- & -/- & -/- & -/- & -/-  & traces of Quartz on XRD, Chromite, CSH \\
	\midrule[1pt]
	SWIR165-short & XRD/SEM & XRD/SEM & XRD/SEM & XRD/SEM & - & -/- & -/- & -/- & -/- & XRD/SEM & -/- & -/SEM\footnote{Mg-rich} & -/- & -/- & -/-  &  well-shaped Mg-rich Mg-carbonate rhombohedra\\
	\midrule
	SWIR165-medium & XRD/SEM & XRD/SEM & XRD/SEM & XRD/SEM & - & -/- & -/- & -/- & -/- & XRD/- & -/- & -/- & XRD/SEM\footnote{Mg-depleted}  & -/- & -/- & Well shaped dolomite rhombohedra \\
	\midrule
	SWIR165-long & XRD/- & XRD/SEM & XRD/SEM & XRD/SEM & - & -/- & -/- & -/- & -/SEM(?) & -/- & -/- & XRD/SEM\footnote{Mg-rich} & XRD/SEM & -/- & -/- &   large euhedral dolomite and calcite crystals\\
	\midrule[1pt]
	SWIR-Agox & -/SEM & -/- & -/- & XRD/SEM & - & XRD/SEM & -/- & -/- & XRD/SEM & -/- & XRD/- & -/- & XRD/SEM & -/- & -/-  & Silver\\
	\midrule
	OPD-Agox & -/- & -/- & -/- & XRD/SEM & traces & -/- & -/- & -/- & XRD/- & -/- & XRD/SEM & -/- & -/- & -/- & -/- & Silver \\
	\midrule[1pt]
	SWIR-Naox & -/- & -/- & XRD/- & XRD/SEM & major & -/- & XRD(?)/SEM & XRD/SEM(?) & -/- & -/- & -/- & XRD/SEM & -/- & XRD/- & -/-  & Natroxalate, Mg-Arfvedsonite \\
	\midrule
	ODP-Naox & -/- & -/- & -/- & XRD/SEM & major & -/SEM & -/- & -/- & -/- & -/- & -/- & -/- & -/- & XRD/SEM & -/- & Albite, Ni-metal, Quartz, chromite, Siderite \\
	\bottomrule
	
	\caption{\label{tab:analyses_exp} Different identified phases with SEM and XRD for each experiments. NA means this technique was not used for the identification. (?) means that the phase was not identified with complete certainty.}	
\end{longtable}

\end{tiny}
\end{landscape}

\subsection{Numerical simulations with PHREEQC}
Phase diagrams were obtained with the free software PhreePlot, based on PHREEQC \citep{Appelo2005,Parkhurst2013a}. Calculations were performed at 300\textcelsius\ and 500 bars with a base electrolyte of 100mM of NaCl in pure water. Chemical activities of the different species of interest were adjusted by the addition of different chloride salts to the solution (\ce{CaCl2}, \ce{MgCl2}, \ce{NaCl}). Silica activity was controlled by silicic acid \ce{H4SiO4}, while p(\ce{CO2}) was controlled by \ce{CO2}(g). The considered mineral phases were pure phases and their thermodynamical properties obtained from the Thermoddem database (\citet{Blanc2012}, https://thermoddem.brgm.fr). 

In many experiments, owing to the high Na activity due to the addition of \ce{NaHCO3} and \ce{Na2C2O4}, the carbonate eitelite \ce{Na2Mg(CO3)2} is detected. Unfortunately, this rare mineral is not included in the Thermoddem database and the only data in the literature is the Gibbs free energy of formation of the mineral from the experimental study of \citet{Konigsberger1992} who measured a value at 25\textcelsius, 1 bar of -2079.76$\pm$1.2 kJ/mol. With no other data to speak of, the only way to incorporate eitelite in the simulations was to reconstruct its thermodynamic data from the sum of its polyhedral contribution following the method of \citet{LaIglesia1994}. This method allows the calculation of the Gibbs free energy of formation as a combination of the Gibbs free energy of formation of the polyhedral units \ce{Na2O}, \ce{MgO} and 2\ce{CO2} using the values from Table \ref{tab:polyhedral_data}. The calculated value at 298 K, 1 bar is -2157.0 kJ/mol, within 4\% of the experimental value. In order to obtain the equilibrium constant as a function of temperature for the dissolution reaction of eitelite \ce{Na2Mg(CO3)2 + 2H+ = 2Na+ + Mg^{2+} + 2HCO3^-}, we used the calculated value from the polyhedral decomposition with the apparent thermodynamic properties $\Delta G_{app}$ of the different dissolved species (\ce{Mg^{2+}}, \ce{Na+} and \ce{HCO3^-}) at different temperatures and at the working pressure (500 bars), as calculated by the code Thermo-ZNS \citep{Lassin2003,Lassin2005}. The Gibbs free energy of the reaction is then obtained with:

	\begin{multline}
		\Delta_r G^0(T,P) = \Delta G_{app}(\ce{Mg^{2+}})(T,P) + 2\Delta G_{app}(\ce{Na+})(T,P) + 2\Delta G_{app}(\ce{HCO3^-})(T,P) - \Delta G_f(eitelite)(T,P)\\  
		+ (T-T_0) \times \left( 2S_0(\ce{Na}) + S_0(\ce{Mg})+3S_0(\ce{O2(g)})+2S_0\ce{C(s)} \right) \\
		+ \left(T\ln(T/T_0)-(T-T_0) \right)\Delta a + (T-T_0)^2 \frac{\Delta b}{2} + \left( \frac{1}{2T} + \frac{1}{2T_0^2} - \frac{1}{T} \right) \Delta c
	\end{multline}
with $\Delta x = 2x(Na)+x(Mg)+3x(O_2(g))+2x(C(s))$ (x = a,b,c, the Maier-Kelley coefficient for the specific heat capacity) and $S_0$ the standard entropy of the constituting elements (Na, Mg, O$2$(g) and C -- \citet{Cox1989}). The molar volume of eitelite is 70.2 cm$^3$/mol \citep{Pabst1973}). 

The calculated equilibrium constant $\ln K = - \frac{\Delta_r G^0}{RT}$ is then fitted according to the classic equation $\log_{10} K(T) = A + B\cdot T + C\cdot T^{-1} + D\cdot \log_{10}(T) + E\cdot T^{-2}$ (adjusted r$^2$=0.99851) and the resulting parameters are included in the PHREEQC database and reported on Table \ref{tab:polyhedral_data}. 

\begin{table}
	\centering 
	\begin{tabular}{cc}
	\toprule
	Polyhedral unit & $g_{i,T}$ kJ/mol \\
	\midrule 
	\ce{Na2O} & -658.4+0.17T \\
	\midrule
	\ce{MgO} & -640.5+0.17T \\
	\midrule
	\ce{CO2} & -472.5+0.11T \\
	\bottomrule
	\end{tabular}
	
	\begin{tabular}{ccccc}
	\toprule 
	A & B & C & D & E \\
	\midrule
	-5927.6 & -0.78426 & 361132 & 2110.1 & -22916016 \\
	\bottomrule
	
	\end{tabular}
		\caption{\label{tab:polyhedral_data} Top: Thermodynamic data for the polyhedral reconstruction \citep{LaIglesia1994}  \\
		Bottom: Results from the fit of the equilibrium constant of eitelite}
\end{table}

\section{Results}

\subsection{Experimental assemblage}

A summary of the observations for the different experiments is reported in Table \ref{tab:analyses_exp}.

\subsubsection{Dry \ce{CO2} experiments}
For both SWIR300g-long and SWIR300g-short experiments, no difference could be seen on the XRD diffractograms with respect to the initial protolith. This total absence of reaction was somewhat expected as all the reactions occurring in the other experiments are mediated by the presence of water and are occurring through a coupled dissolution/precipitation mechanism. The absence of water prevents the dissolution of the minerals and thus the reprecipitation of the new phases. This is perfectly consistent with the study of dry carbonation of serpentine and other ultramafic minerals in the literature where it was shown that a pretreatment of the feedstock to activate the MgO fraction was necessary before carbonation \citep{Zevenhoven2002,Zevenhoven2016}

\subsubsection{Runs at 300\textcelsius}

\begin{landscape}
	\begin{figure}
		\centering \includegraphics[width=\textwidth]{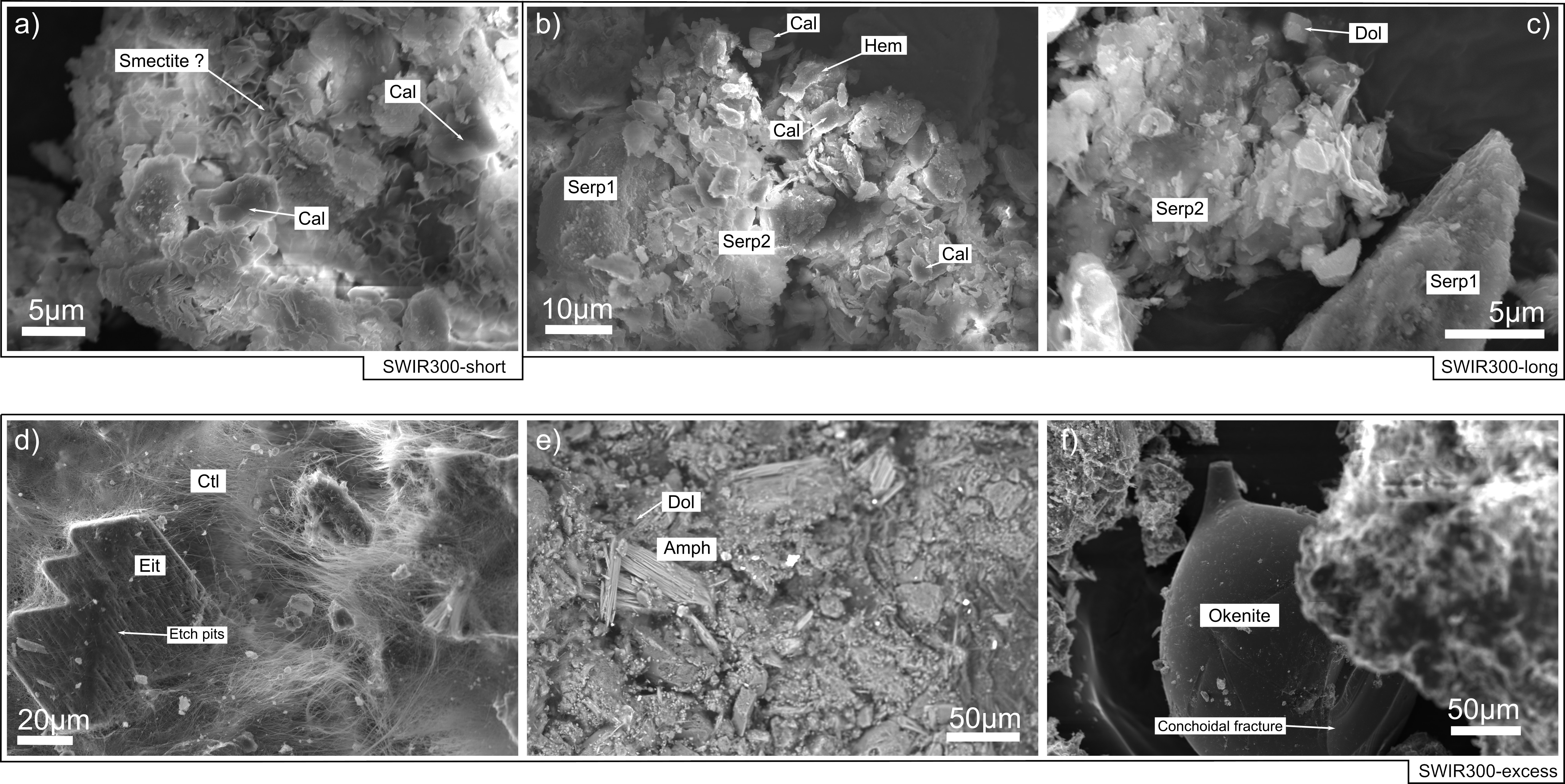}
		\caption{\label{fig:SWIR300}SEM pictures of experiments SWIR300-medium (a), SWIR300-long (b,c) and SWIR300-excess (d,e,f). Note the particular structure on (a) which could be associated with clay minerals (smectite ?). Figures (b) and (c) show similar grains of initial serpentine (Serp1) covered with more flaky serpentine (Serp2) potentially from secondary serpentine precipitation after olivine and pyroxene alteration. A grain of hematite (Hem) is identified on image (b), while image (c) shows a slightly Mg-depleted, euhedral rhombohedral dolomite crystal (Dol). Figure (d) shows the very extensive fibrous precipitation of chrysotile (Ctl) along with large grains of eitelite (Eit). This particular grain presents visible dissolution figures (etch pits). Image (e) presents some serpentine and fibrous amphibole (Amph) as well as a small dolomite crystal. Figure (f) shows an interesting mineral, identified as the CSH (hydrous calcium silicate, typical of cement minerals) okenite due to its particular fracture pattern.}
	\end{figure}
\end{landscape}

	\paragraph{SWIR300 time series}
The evolution of the mineralogy at 300\textcelsius\ and 500 bars exhibits a quick disappearance of the primary minerals (pyroxenes, olivine and aragonite) replaced by a mixture of carbonates and serpentine. After 7 days of reaction, both clino- and orthopyroxenes have completely disappeared from the XRD diffractogram while a trace amount of forsterite remains. In the meantime, the serpentine signal increased significanrly. Carbonates are composed principally of Mg-calcite ($\approx$5\%wt MgO). On SEM images, calcite crystals appear small and anhedral (Fig. \ref{fig:SWIR300}a and b). This precipitation of calcite is clearly associated with the recrystallization of aragonite from the starting material. This recrystallization occurs rapidly as no aragonite can be detected already in the shortest experiment. In parallel of aragonite/calcite recrystallization, a peak corresponding to a dolomite composition is visible in the XRD diffractogram for the longest experiment. The position of this XRD peaks on the diffractogram are slightly shifted towards larger d (smaller 2$\theta$) suggesting a departure from the ideal dolomite composition due to either an excess of \ce{Ca} or a substitution of Fe$^{2+}$ for Mg. SEM/EDS analysis of these crystals (Fig. \ref{fig:SWIR300}c) support the former option as the dolomite crystals contain little to no Fe and a slightly Mg-depleted composition. 

In terms of redox conditions, a large amount of iron is oxidized to Fe$^{3+}$ as evidenced by the brick-red color of the powders and the hematite peaks detected on each diffractogram. Iron is also contained in the secondary serpentine up to 10-12\%wt (Figs. \ref{fig:SWIR300}c and d). While it is is not possible to determine if this iron is in the +II or the +III form, the amount is larger than in the initial serpentine and suggests a precipitation of iron-rich serpentine from the dissolution of olivine and pyroxenes. This additional precipitation is visible on SEM images showing grains completely covered with serpentine flakes contrasting with the aspect of the grains from the starting material (Figs. \ref{fig:SWIR300}b and c). The participation of the initial serpentine (i.e. the dissolution of the starting serpentine and its reprecipitation) cannot however be assessed. In terms of serpentine polymorph, no fiber could be found and the flaky aspect on SEM images suggests the occurrence of lizardite only. 

Finally, a clay phase is detectable on XRD with a large peak around the 10\AA mark and is a probable sink for the Al released by the dissolution of pyroxenes. This clay phase is however not unambiguously identified with SEM, but could be associated with the net-like structure visible on Figure \ref{fig:SWIR300}a. Unfortunately, due to the small size of the minerals and their orientation, no clear composition and thus no confirmation could be obtained. 

\paragraph{OPD300} The effect of carbonation on the ODP serpentinite is on the contrary very limited. No carbonate of any sort can be detected neither on SEM images nor on XRD. This indicates that carbonation of serpentine at 300\textcelsius\ and 500 bars is not happening within this time frame despite the thermodynamical drive, probably due to the very slow dissolution rate of serpentine and the even slower precipitation rate of magnesite under these conditions \citep{Saldi2012}. Minor chromite, iron oxides and Nickel sulfures can still be observed highlighting the little effect of the treatment on the overall composition. Importantly, the color of the resulting powder remained green: despite a redox drive towards iron oxidation as highlighted by the behavior of the SWIR serpentinite, no hematite was produced because of the absence of serpentine dissolution and thus the absence of Fe release. 

\paragraph{SWIR300-excess} Due to the finite solubility of sodium bicarbonate in water, the actual quantity of \ce{CO2} in each SWIR300 experiment is between 1/3 and 1/2 of the total divalent cations from olivine and pyroxenes (Table \ref{tab:summary_exp}). In order to force the reaction of the serpentine, it was necessary to add enough bicarbonate to not only have enough for the carbonation of both the primary minerals and serpentine, but also to increase \ce{CO2} activity. To do so, the usual SWIR300 was prepared, and then extra bicarbonate powder was added on top to reach a total of \ce{CO2}/cations = 3.5.  

The striking observation on SEM images from SWIR300-excess is the important amount of thin ($\approx$200nm) chrysotile fibers about 20$\mu$m long creating a weaving in which the other minerals are embedded (Fig. \ref{fig:SWIR300}d). This suggests a very large precipitation of chrysotile serpentine, very much in contrast with the previous experiments. Additionally, some occurrences of sodic amphibole reinforce the observation of a much more important alteration of the starting powder than in the previous SWIR experiments. The sodic amphibole presents a composition related to Riebeckite $\ce{Na2(Fe^{2+}Mg^{2+})_3Fe^{3+}2Si8O22(OH)2}$ and Mg-Arfvedsonite $\ce{NaNa2(Mg^{2+}_4Fe^{3+})Si8O22(OH)2}$. The lack of precision of EDS on powders unfortunately precludes any more precise determination. 

Carbonates are principally found as very large (100$\mu$m) eitelite \ce{Na2Mg(CO3)2} crystals. XRD diffractograms also reveals, similarly to the other SWIR300 experiments, the presence of dolomite with peaks shifted towards smaller 2$\theta$. The analysis of a the dolomite crystal on Figure \ref{fig:SWIR300}e shows a small amount of iron in the composition as well as a small Mg-depletion. Interestingly, eitelite crystals also exhibit some dissolution features and etch pits (Fig. \ref{fig:SWIR300}d) suggesting that eitelite might be a temporary, metastable mineral in these conditions, dissolving in favor of the more stable dolomite. This is in accordance with the presence of dolomite at the end of the time series at 300\textcelsius. The very high Na-activity associated with the extra \ce{NaHCO3} created a secondary reaction path through eitelite instead of the reprecipitation of dolomite from calcite, actually skipping calcite entirely. The consequence of the precipitation of eitelite instead of calcite is that the calcium has then to be incorporated in another mineral phases. This phase is not visible on XRD, but can be seen on some SEM images such as Figure \ref{fig:SWIR300}f. The composition of the mineral visible on Figure \ref{fig:SWIR300}f is pure Ca-Si and  correspond to CSH composition (water is not quantified on EDS). The XRD diffractograms does not allow its identification but it could be related to hillebrandite (\ce{Ca6Si3O9(OH)6}) or more likely okenite (\ce{Ca3[Si6O15]*6(H2O)}) due to its rounded shape and the conchoidal fracture. 

In terms of mass balance, the strong precipitation of chrysotile associated with the sodic amphibole and the CSH minerals implies a high Si activity consistent with an alteration of both the primary minerals and the initial serpentine itself, but however not high enough for talc precipitation. Finally, similarly to the time series, the XRD presents a clay peak around 10\AA, which is the probable sink for Al released during the dissolution of the pyroxenes. 

\subsubsection{165\textcelsius\ time series}
\begin{figure}
	\centering \includegraphics[width=\textwidth]{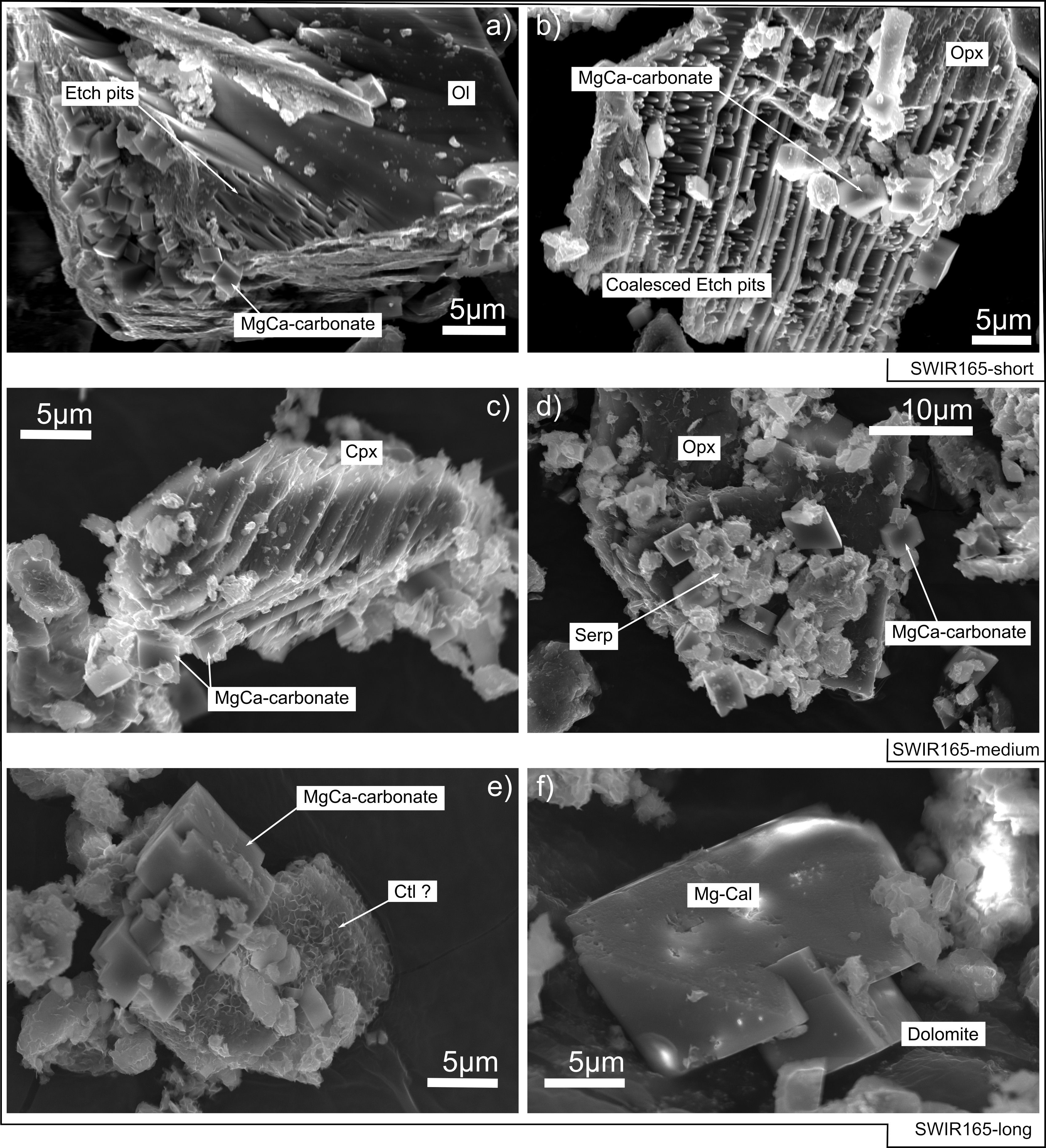}
	\caption{\label{fig:SWIR165} SEM figures of experiments SWIR165. Figure (a) shows an olivine grain with visible etch pits and small euhedral carbonates. Figure (b) presents an orthopyroxene with alteration features along the cleavages as well as coalesced etch pits. Image (c) depicts a very alterered clinopryxoxene with visible dissolution features following the cleavage planes. Figure (d) presents an orthopryxene crystal covered with neoformed serpentine and MgCa-carbonate rhombohedra. It is impossible from the pictures to assess if it is a Mg-rich calcite or a Mg-depleted dolomite. Figure (e) presents a serpentine grain associated with a large crystal of carbonate and some neoformed serpentine flakes. The surface of the grain is also covered by a net-like structure which could be the onset of chrysotile precipitation. Finally figure (f) shows a Mg-rich calcite mineral associated with a dolomite crystal. Note that in both figures from the longer experiments (e and f) the carbonate crystals are significantly larger than for the shorter experiment.  }
\end{figure}

Kinetics of reaction at 165\textcelsius\ are drastically slower, leading to a much slower dissolution of the primary minerals and a slower precipitation of the secondary ones. After 63 days of reaction (SWIR165-long), initial pyroxenes and olivine are still present and visible both with XRD and SEM. At 13 and 21 days (respectively SWIR165-short and SWIR165-medium), the pyroxene and olivine crystals present visible dissolution features with dissolution along the cleavage plans and etch pits (e.g. Fig. \ref{fig:SWIR165}a, b, c). Similarly, aragonite, which disappeared quickly at 300\textcelsius\, remains visible in XRD diffractograms after 21 days of reaction, albeit in very small quantities. A crystal of aragonite was also identified on SEM images for the 13 days experiment. Aragonite eventually disappears on the 63 days-long experiment. Carbonate precipitation is in the meantime already visible on SEM images after 13 days, with the presence of large euhedral rhombohedral minerals ($\approx 5\mu m$ large) containing up to 40\% MgO after 21 days. These carbonate corresponds on XRD to the same shifted dolomite peak as for experiments SWIR300 and SWIR300-excess, indicating that the minerals are already assembling into the Mg-depleted dolomite structure observed at 300\textcelsius. After 63 days, carbonates get even more enriched in Mg with a composition approaching the ideal dolomite composition. In parallel however, a pure calcite peak, which was not present in the previous experiments is also visible and can be related to the more Mg-poor carbonate visible on Figure \ref{fig:SWIR165}f. This occurrence of calcite at the end of the experiment is surprising as it was expected that the longer the experiment, the more Mg is being released and thus the more Mg-enriched carbonates become. One explanation could be that the rate of release of Mg is slowing down due to the low reactivity of the primary minerals and the inability to alter serpentine, while the rate of calcium remains high with the continuous dissolution of aragonite. As a result, the Mg/Ca ratio actually decreases leading to the precipitation of Mg-calcite instead of dolomite.

The initial grains on SEM images appear covered with new flakes of serpentine, slightly richer in Fe than the initial serpentine, similarly to SWIR300. A net-like structure is also clearly visible after 21 days and further (Fig. \ref{fig:SWIR165}e) and the absence of the clay peak around 10\AA\ could indicate that this structure is the onset of the precipitation of fibrous serpentine (similar to images from \citep{Lafay2018}). Unfortunately, due to the small size of the feature, no clear composition could be retrieved from the SEM images. 

Finally, the absence of any iron oxide on both SEM images and XRD suggests that all the iron is incorporated in the serpentine while carbonates are virtually iron-free. This is consistent with studies of serpentinization showing that iron is more easily incorporated in the serpentine structure at lower temperatures \citep{Klein2009}. The color of the powders goes however from yellow for the short and medium experiments to light red after 63 days implying nevertheless some hematite precipitation.

\subsubsection{Silver Oxalate runs}
\begin{figure}
	\centering \includegraphics[width=\textwidth]{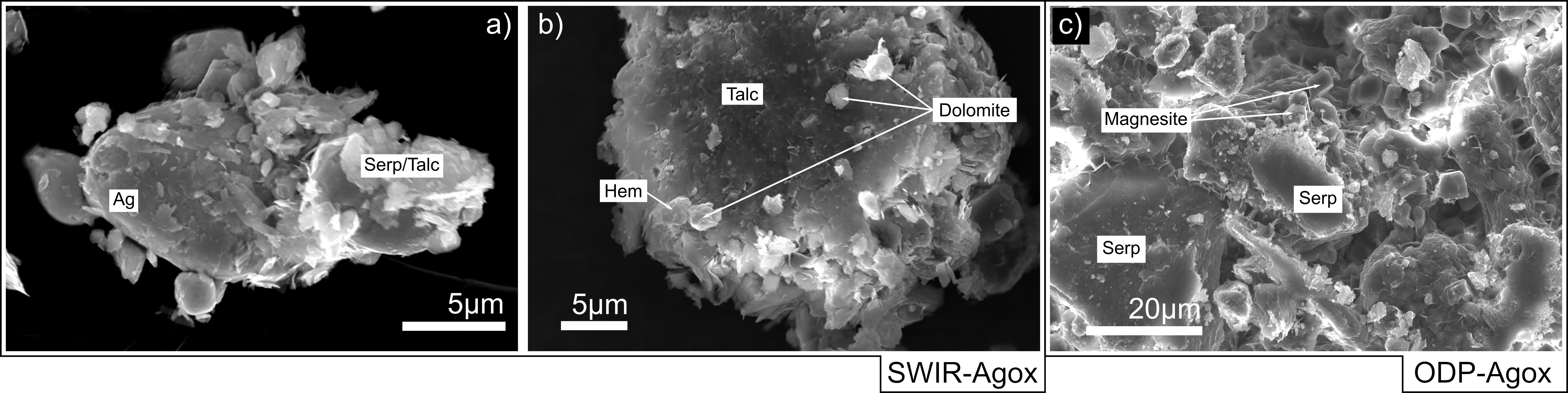}
	\caption{\label{fig:Agox} SEM images of the silver oxalate runs (a) and (b) from SWIR-Agox and (c) from ODP-Agox. Image (a) represents metallic silver covered with talc/serpentine sheets showing that the precipitation of these minerals occurred after the thermal decomposition of oxalate and are thus a result of the carbonation. Figure (b) shows a grain covered with talc and a few anhedral crystals of dolomite as well as a crystal of iron oxide, identified as hematite from the XRD diffractograms. Finally image (c) shows a mixture of serpentine with large euhedral crystals of magnesite. The embedding of the magnesite crystals in the serpentine highlights a co-precipitation of these two minerals and thus a precipitation of a neoformed serpentine during the experiment.}
\end{figure}

We observed that with sodium bicarbonate, serpentine is only destabilized after all the primary minerals are consumed and if the \ce{CO2}activity is high enough. To further study the influence of the \ce{CO2} activity on the reaction, we used silver oxalate \ce{Ag2C2O4} which thermally decomposes into 2Ag(s) and 2\ce{CO2}, effectively providing a much higher \ce{CO2} activity than the maximum solubility of bicarbonate ($m_{ce{CO2}}/(m_{ce{CO2}}+m_{ce{H2O}}) = 0.5$, with $m$ the mole quantity of water and \ce{CO2}). The resulting assemblage from the SWIR serpentinite after 30 days of reaction at 300\textcelsius\ presents a mixture of serpentine associated with talc, forming large sheets, 10 $\mu$m large with no occurrences of fibrous serpentine (Fig. \ref{fig:Agox}a). Carbonate are present as anhedral iron-free dolomite, with composition close to the stoichiometric dolomite, and visible on both SEM (Fig. \ref{fig:Agox}b) and XRD. In addition, some magnesite peaks are visible on the XRD but magnesite could not be identified unambiguously on SEM images. Since the dolomite is iron-free, the iron is partitionned between hematite (visible on XRD and confirmed also by the red color of the resulting powder), talc and neoformed serpentine. The complete absence of calcite shows that the alteration liberated enough Mg to recombine with all the Ca liberated by the aragonite dissolution and form dolomite, while the surplus became magnesite. Accordingly, this suggests a higher degree of alteration of the primary minerals and thus a likely involvement of serpentine in the overall mass balance. The presence of talc and absence of quartz reveals however that the alteration is not complete and a significant amount of Mg is still present in the silicates minerals. 

This reactivity of serpentine to higher \ce{CO2} activity through silver oxalate addition is confirmed by the results from the run with the ODP serpentinite. Indeed, in opposition to the results of ODP300, the serpentine minerals present some obvious dissolution features shown on Figure \ref{fig:Agox}c. While serpentine is still the main constituent of the assemblage after 30 days, euhedral crystals of magnesite about 5$\mu$m large can also easily be seen on SEM images. These  crystals contain at most a few percent of Fe. The absence of any Si-rich phase (talc, quartz\ldots) suggests that a new serpentine phase is precipitating to incorporate the Si released by the initial serpentine dissolution. Finally, similarly to the SWIR run, the ODP run contains hematite visible on both XRD and SEM and confirmed by the characteristic red color of the resulting powder. 

\subsubsection{Sodium Oxalate Runs}
All presented experiments were proceeded in rather oxidative conditions as evidenced by the brick-red color of the resulting powders as well as hematite being present in almost all runs except for ODP300 where the lack of reactivity of the serpentine prevented any oxidation of iron and SWIR300-excess where the iron was incorporated in the amphibole phase and thus not oxidized. In order to present similar conditions of \ce{CO2} activity to the silver oxalate runs, while providing a more reducing environment, sodium oxalate \ce{Na2C2O4} was used. Indeed, sodium oxalate decomposes with temperature as: $\ce{Na2C2O4 = 2Na^+ + CO_3^{2-} + CO}$. The carbon monoxide gets eventually oxidized in \ce{CO2} providing the overall same amount of \ce{CO2} activity to the system but with much more reducing conditions. This is evidenced by the presence of magnetite instead of hematite. The other consequence of using sodium oxalate is the very high activity of sodium, similar to the activity achieved when using excess sodium bicarbonate.  

Reaction of SWIR serpentinite with sodium oxalate leads to the disappearance of olivine and orthopyroxene  while clinopyroxene remains visible in the XRD diffractogram but is not found in SEM images. Eitelite dominates the carbonate phase similarly to the SWIR300-excess run, with a secondary Mg-poor calcite. Interestingly, no dolomite is detected and the reprecipitation of eitelite did not occur. The much larger Na activity compared with SWIR300-excess probably stabilizes eitelite and prevents its reprecipitation as dolomite, explaining the presence of the calcite as the Ca sink. Interestingly, no CSH is observed contrary to SWIR300-excess suggesting a smaller Si/\ce{CO2} ratio. The presence of a much larger amount of amphiboles suggests that the \ce{CO2} activity was too high for CSH precipitation and that the released Si was then incorporated into the amphibole phase. This amphibole phase shows a composition very close to the amphiboles from SWIR300-excess. Also, similarly to SWIR300-excess, the secondary serpentine is precipitating as the chrysotile polytype. However, they appear as sheaf of thicker fibers, 40-50$\mu$m long, and much straighter than the fibers in SWIR300-excess. 

Finally, large grains of natroxalate (i.e. the initial sodium oxalate) are both visible on SEM and XRD. This means that the thermal decomposition of the initial powder was not complete during the experiments leading to lower \ce{CO2} activity compared with the silver oxalate runs where all the \ce{CO2} was released. 

Sodium oxalate runs with the ODP serpentinite is presenting very large grains (up to 200$\mu$m) of euhedral eitelite indicating a strong alteration of the initial serpentine. These eitelite crystals also do not present any dissolution features similarly to SWIR-Naox and in contrast with SWIR300-excess. Alteration of the primary serpentine is very advanced leading to the formation of talc as well as large quartz grains up to 100$\mu$m but with no amphibole. The advanced alteration of the serpentinite is also visible through the observation of well-shaped columnar albite as sinks of the aluminum from the serpentine (despite a small initial Al content of about 1\%wt). The strongly reducing conditions and the strong alteration of the serpentine are also underlined by the detection of metallic Nickel reduced from the Ni$^{2+}$ originating from the dissolution of the initial serpentine. Interestingly, the initial serpentine was also replaced by a new fibrous serpentine, visible on all SEM images, with a structure similar to SWIR300-excess. Finally, in constrast with all other experiments, iron is incorporated in a siderite phase. 

\begin{landscape}
\begin{figure}
	\centering \includegraphics[width=\textwidth]{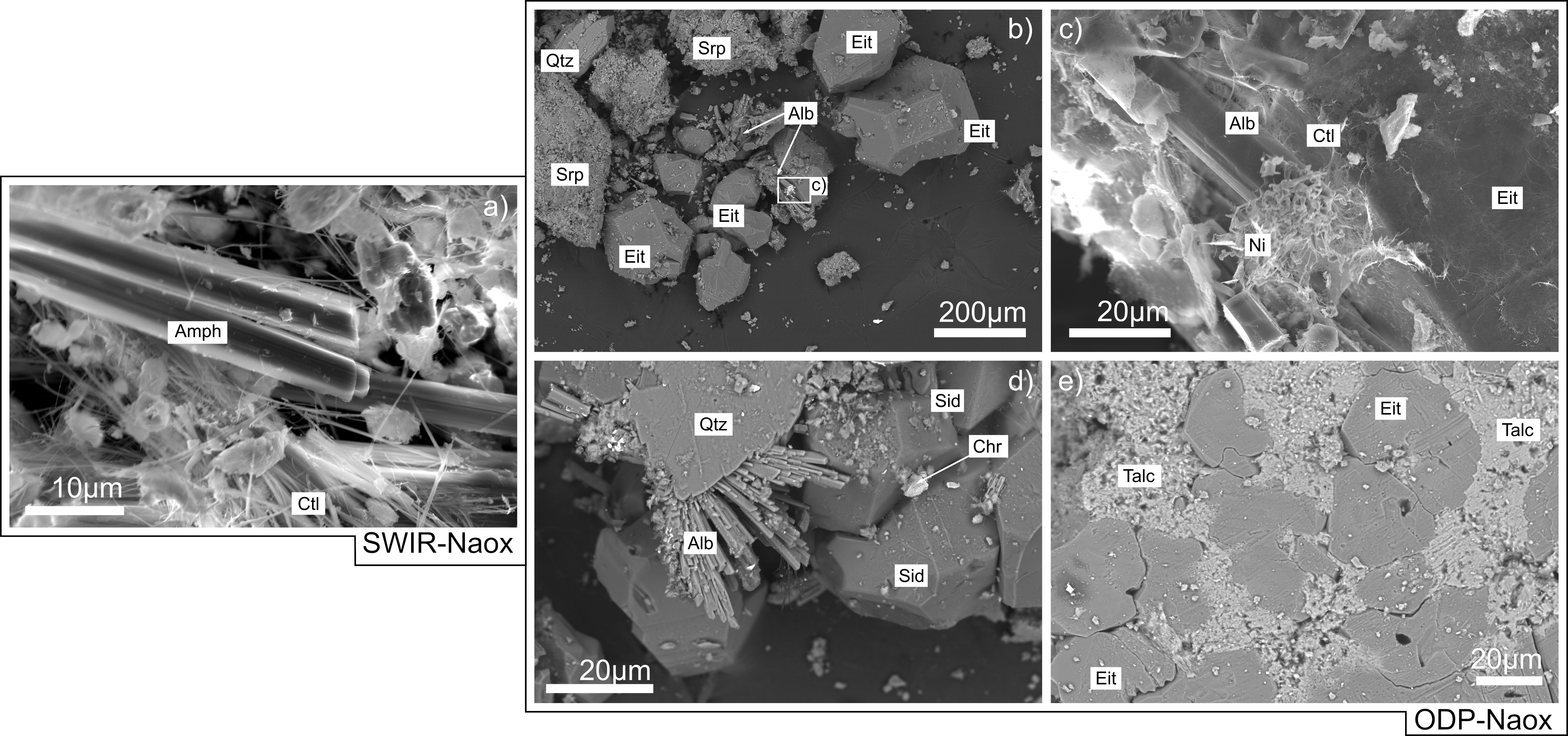}
	\caption{\label{fig:Naox} SEM images of the sodium oxalate runs. Figure (a) is from SWIR-Naox and shows a large fibrous sodic amphibole (Amph) as well as numerous thin chrysotile fibers (Ctl). Figure (b) to (d) are from ODP-Naox. Image (b) shows well-shaped euhedral eitelite (Eit), serpentine (Srp) and quartz (Qtz) grains. Note the large size of the different grains compared to the other experiments. Figure (c) is a close-up from figure (b) showing solid Ni on top of an Albite (Alb) pillar and covered with very thin Chrysotile fibers. Image (d) shows more albite pillars next to quartz and siderite (Sid) grains and a small chromite (Chr) fragment. Finally, image (e) shows large eitelite crystals embedded in a talc matrix showing a simultaneous precipitation of both minerals.}
\end{figure}
\end{landscape}

\section{Discussion}

The experiments presented aboce allow to draw an accurate picture of the behavior of natural serpentinites subjected to carbonate-rich waters. The first minerals to react are the primary ultramafic minerals, olivine and pyroxenes with orthopyroxenes being the fastest as reported in previous studies on peridotite carbonation and serpentinization \citep{McCollom2020,Allen2003}. Conversely, serpentine is only reacting if the primary minerals are exhausted and if the \ce{CO2} activity is high enough \citep{Hovelmann2011,Grozeva2017}. That being said, numerical simulations and natural examples in the literature usually present extensive alteration of the serpentine even at lower temperatures and \ce{CO2} activities than in our experiments, suggesting that the absence of alteration of the serpentine in our experiments might be due to the extremely slow kinetic rates of serpentine alteration rather than unfavorable thermodynamics. It is possible that much longer experiments would have seen more alteration of the serpentine (provided there was enough bicarbonate to react with olivine, pyroxenes and serpentine). 

\begin{figure}
	\centering \includegraphics[width=\textwidth]{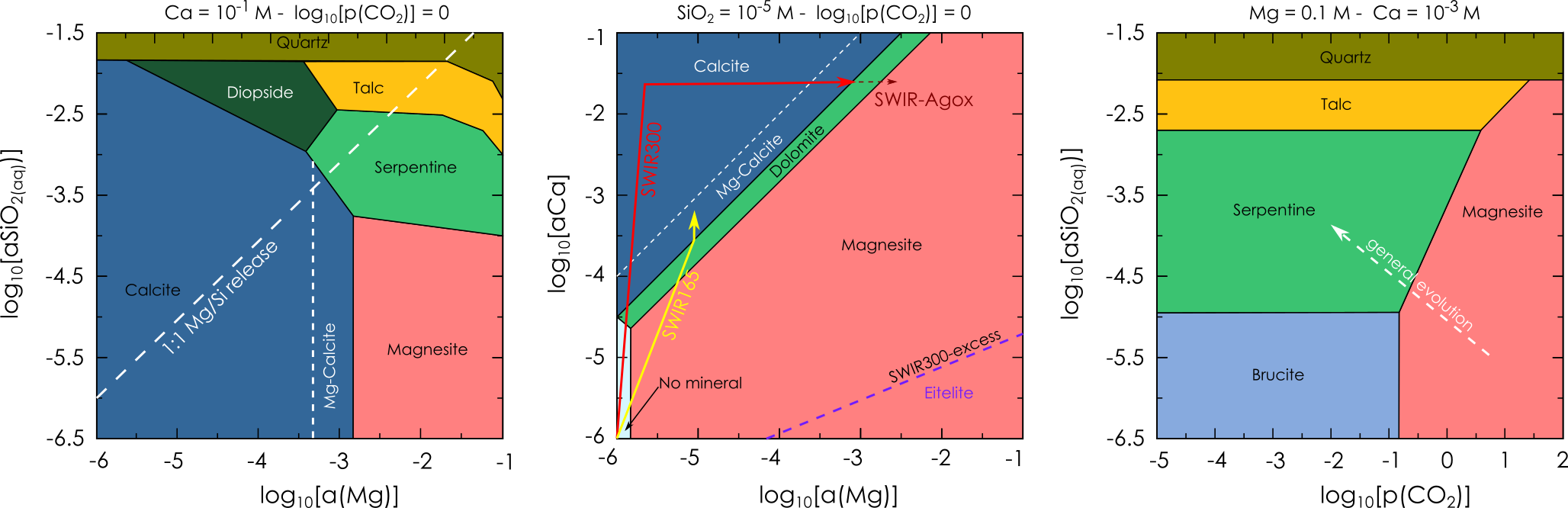}
	\caption{\label{fig:phase_diagrams}Phase diagrams in the system Mg-Ca-CO$_2$-SiO$_2$ plotted with the help of PhreePlot. The first diagram (left panel) represents the major phases as a function of \ce{SiO2} and Mg activity with fixed Ca and p(\ce{CO2}). The second diagram (center panel) represents the major phases as a function of Ca and Mg activities, with fixed Si and p(\ce{CO2}). Finally, the last diagram (right panel) represents the major phases as a function of Si and p(\ce{CO2}) with fixed Mg and Ca. }
\end{figure}

In a very interesting twist, relicts clinopyroxenes were observed on XRD for both SWIR300-excess and SWIR-Naox. We posit that their presence is due to the high Ca and low Mg and \ce{CO2} activities in the fluid during these experiments. A stability domain of diopside is indeed visible on the phase diagram shown on Figure \ref{fig:phase_diagrams}. This is consistent with the precipitation of eitelite, which occurs in this case instead of calcite, and reduces both Mg and \ce{CO2} activities while leaving the calcium in solution as well as silica. The extent of the associated CSH and dolomite precipitation for SWIR300-excess, as well as calcite in SWIR-Naox, is too low to reduce Ca activity and thus leads to the stability of cpx.

Precipitation of carbonates occurs differently depending on the availability of the different cations and results from a subtle balance between the different cations activities as illustrated by the cpx example. In the SWIR300 runs, the instantaneous dissolution of aragonite (relatively to the dissolution of the primary minerals) and its quick recrystallization as calcite leads to a progressive dolomitization of the primary calcite with increasing Mg content in the fluid (red path on Figure \ref{fig:phase_diagrams}). On the contrary, the much slower aragonite recrystallization at 165\textcelsius\ allows for a more equilibrated Mg/Ca fluid composition from the beginning and the precipitation of carbonates with a composition closer to the ideal dolomite without passing by calcite. With increasing time, the Mg/Ca composition of the fluid decreases as the primary minerals dissolution slows down while aragonite keeps releasing Ca leading to the precipitation of a lower Mg-calcite towards the end (Yellow path on Figure \ref{fig:phase_diagrams}). In all case, the combination of Mg-calcite and dolomite is similar to classic experimental and field observations \citep{Ueda2017,Grozeva2017}

For the Na-rich experiments, eitelite \ce{Na2Mg(CO3)2} is the principal form of carbonate, supplanting calcite and dolomite. However, thermodynamic calculations show that eitelite is systematically unstable with respect to dolomite and magnesite at the SWIR300-excess conditions, and presents  a small stability field if only calcite and eitelite are taken into account. This instability of eitelite is consistent with the dissolution features on the crystals for SWIR300-excess suggesting eitelite is metastable and precipitates because of favorable kinetics with respect to dolomite. However, for both Naox experiments, eitelite appears on the contrary pristine suggesting that the very high Na activity is enough to stabilize it. 

In any case, except for ODP-Agox where no other divalent cation is available and Na activity is too low, magnesium is always incorporated in the carbonate phase through the mediation of another cation, either calcium for Mg-calcite and dolomite, or sodium for eitelite. The preliminary crystallization of an intermediary phase is due to the well-known difficulty of Mg to readily precipitate as magnesite due to its high hydration energy \citep{Saldi2012}. The co-precipitation along with another cation (Na or Ca) creates a kinetically easier pathway through a metastable intermediate. This is of high importance for the carbon dioxide mineralization, for which the slow kinetics of precipitation of magnesite is an issue. The addition of another cation will promote and accelerate the incorporation of Mg into the carbonate phase and thus accelerate the whole carbonation process. 	

The shape of carbonates is also an indication of the reaction kinetics and illustrate clearly the influence of temperature. At 165\textcelsius, the carbonate crystals are well-shaped euhedral rhombohedra (Fig. \ref{fig:SWIR165}), while at 300\textcelsius, the carbonates are completely anhedral and much smaller (Fig. \ref{fig:SWIR300}). The number and shape of crystals is indeed, directly linked to the distance from equilibrium. For close-to-equilibrium situations, nucleation of new crystals is unfavorable and the growth of existing crystals is the process by which supersaturation is consumed. The resulting slow growth rates allows for a proper organization of the atoms on the crystallographic sites and thus the euhedral characteristics at 165\textcelsius. At far-from-equilibrium conditions, the growth of existing crystals is usually not fast enough to consume the supersaturation and additional nuclei are created. The fast precipitation does not allow the same organization as for close-to-equilibrium conditions and leads to the numerous small anhedral crystals. Interestingly, the magnesite crystals for ODP-Agox (Fig. \ref{fig:Agox}c) appear well-shaped despite the higher temperature. This is probably linked to the inherently slow kinetics of magnesite precipitation. 

While carbonates are precipitating, the dissolution of the primary minerals releases also Si in solution. As the Si activity rises, serpentine first precipitates, then talc and then quartz following the increase in Si/Mg ratio, as Mg is also involved in carbonate precipitation. In the experiments, talc is only seen for the SWIR-Agox and ODP-Naox experiments while quartz is only detected in ODP-Naox. This means that for most experiments, serpentine is the main sink for Si and that the activity never gets to the talc/serpentine equilibrium. Besides, no brucite is detected indicating a Si activity high enough to destabilize brucite. This Si activity is due to the dissolution of pyroxenes. Interestingly, the neoformed serpentine polymorph depends on the initial experimental conditions. For most experiments, the observed serpentine is flaky and corresponds to a morphology related to lizardite. However, in three experiments (SWIR300-excess, SWIR-Naox and ODP-Naox), the precipitating serpentine is strongly fibrous chrysotile. These three experiments have in common a very high activity of Na and the presence of eitelite, as well as an absence of iron oxidation and a green color. The role of Fe$^{2+}$/Fe$^{3+}$ in the stabilization of the crystallographic structure of the serpentine could be the reason of these two different assemblages, with Fe$^{3+}$ stabilizing the lizardite structure while Fe$^{2+}$ is usually more associated with chrysotile \citep{Evans2008,Mayhew2020}. Another explanation could come from the very high Na activity in all the experiments where chrysotile is observed but the actual mechanism by which this high Na activity could lead to the formation of chrysotile instead of lizardite is unclear.

Finally, no passivating layer is observed on any experimental run. despite being often described in the literature \citep{Bearat2006,Saldi2015} Silica is always incorporated in a crystalline phase, and does not occur as amorphous silica or Mg-depleted layer on pyroxenes or olivine. Precipitation of sheet silicates is however potentially passivating as on Figure \ref{fig:Agox}a,b but the completion of the reaction for the SWIR-300 experiments as well as the advanced alteration for the Naox and Agox runs suggests that no passivation of the primary minerals occurs and that they are consistently being altered by the reacting fluid until complete disappearance.

\section{Conclusion}
As highlighted in the literature, all ultramafic minerals are not equal when it comes to carbonation. While orthopyroxenes, clinopyroxenes and olivine react readily and quickly to carbonated fluids, serpentine is much slower to react and requires high activity of \ce{CO2}. Experiments at 165\textcelsius\ and 300\textcelsius\ have shown that a solution of 2\%wt NaCl and 5\%wt \ce{NaHCO3} is insufficient to destabilize serpentine within 30 days at 300\textcelsius\ and 60 days at 165\textcelsius, but can lead to extensive carbonation of the primary minerals (ol, opx and cpx). On the contrary, the use of silver and sodium oxalate, or the use of excess bicarbonates generate much more aggressive conditions leading to the destabilization of the initial serpentine. The use of natural serpentinites for carbon mineralization is then possible if enough \ce{CO2} is provided to the system and given sufficient time for the reaction. Interestingly, the addition of other cations such as Na or Ca, which do not have the same kinetic limitations for the carbonate precipitation enhances the Mg uptake in the carbonate phase and accelerates the overall carbonation. Experiments also highlight the potential complications when trying to use carbonation as a way for asbestos inertization. Indeed, some experiments, even with very high \ce{CO2} activities, precipitate a significant amount of asbestiform materials (amphibole included), sometimes through the alteration of non-asbestiform serpentine. The process needs then to be carefully controlled in order to not make the matter worse. 

Finally, despite the very specific solution and somewhat "unnatural" composition of the fluids, some parallels between the experiments and natural settings such as listwanite formation can be drawn. In particular, experiments demonstrate that the carbonation of serpentine in natural settings is most likely to occur only after all the olivine and pyroxenes have been carbonated, as the \ce{CO2} activities are unlikely to be high enough to alter both serpentine and primary minerals at once. Moreover, the alteration and carbonate formation is not linear and depends strongly on the kinetics of alteration as well as on the composition of the percolating fluids. Depending on the initial Ca content and on the temperature, dolomite precipitation for example might follow calcite precipitation or the other way around. However, the comparison of the presented experiments with natural examples of carbonation needs to be done carefully as the coupling between chemistry, hydrodynamics and sometimes mechanics plays a fundamental role in the final assemblage, which cannot be reproduced with simple batch experiments \citep{Osselin2016}. 

\section*{Aknowledgments}
We would like to thank P. Penhoud for the XRD data. This research was supported by the LABEX Voltaire (ANR-10-LABX-100-01) and EQUIPEX PLANET (ANR-11-EQPX-0036).

\bibliographystyle{apalike}
\bibliography{biblio}

\end{document}